\documentclass[iop,apj]{emulateapj}
\usepackage{natbib}
\usepackage[dvips]{color}
\usepackage{multirow}
\usepackage{amsmath}
\usepackage{amsfonts}
\newcommand{\mum}{$\mu$m}
\newcommand{\herschel}{{\it Herschel}}

\newcommand{\kms}{km\,s$^{-1}$}
\newcommand{\midco}{CO\,(6$-$5)}
\begin{document}
\shorttitle{ALMA \midco\ emission in NGC7130}
\shortauthors{Zhao et al.}
\title{ALMA Imaging of the \midco\ Line Emission in NGC\,7130$^\star$}

\author{Yinghe Zhao\altaffilmark{1,2,3,4}, Nanyao Lu\altaffilmark{5,6,7}, C. Kevin Xu\altaffilmark{7}, Yu Gao\altaffilmark{3,4}, Loreto Barcos-Mun\~{o}z\altaffilmark{8}, Tanio D\'{i}az-Santos\altaffilmark{9}, Philip Appleton\altaffilmark{7}, Vassilis Charmandaris\altaffilmark{10,11}, Lee Armus\altaffilmark{12}, Paul van der Werf\altaffilmark{13}, Aaron Evans\altaffilmark{8, 2}, Chen Cao\altaffilmark{14, 15}, Hanae Inami\altaffilmark{16}, and Eric Murphy\altaffilmark{7}}
\altaffiltext{$\star$}{The National Radio Astronomy Observatory is a facility of the National Science Foundation operated under cooperative agreement by Associated Universities, Inc.}
\altaffiltext{1}{Yunnan Observatories, Chinese Academy of Sciences, Kunming 650011, China; zhaoyinghe@gmail.com}
\altaffiltext{2}{National Radio Astronomy Observatory, Charlottesville, VA 22904, USA}
\altaffiltext{3}{Purple Mountain Observatory, Chinese Academy of Sciences, Nanjing 210008, China}
\altaffiltext{4}{Key Laboratory of Radio Astronomy, Chinese Academy of Sciences, Nanjing 210008, China}
\altaffiltext{5}{National Astronomical Observatories of China, Chinese Academy of Sciences, Beijing 100012, China}
\altaffiltext{6}{China-Chile Joint Center for Astronomy, Chinese Academy of Sciences, Camino El Observatorio, 1515 Las Condes, Santiago, Chile}
\altaffiltext{7}{Infrared Processing and Analysis Center, California Institute of Technology 100-22, Pasadena, CA 91125, USA}
\altaffiltext{8}{Department of Astronomy, University of Virginia, 530 McCormick Road, Charlottesville, VA 22904, USA}
\altaffiltext{9}{N\'{u}cleo de Astronom\'{i}a de la Facultad de Ingenier\'{i}a, Universidad Diego Portales, Av. Ej\'{e}rcito Libertador 441, Santiago, Chile}
\altaffiltext{10}{Department of Physics, University of Crete, GR-71003 Heraklion, Greece}
\altaffiltext{11}{IAASARS, National Observatory of Athens, GR-15236, Penteli, Greece}
\altaffiltext{12}{Spitzer Science Center, California Institute of Technology, MS 220-6, Pasadena, CA 91125, USA}
\altaffiltext{13}{Leiden Observatory, Leiden University, P.O. Box 9513, 2300 RA Leiden, The Netherlands}
\altaffiltext{14}{School of Space Science and Physics, Shandong University at Weihai, Weihai, Shandong 264209, China}
\altaffiltext{15}{Shandong Provincial Key Laboratory of Optical Astronomy and Solar-Terrestrial Environment, Weihai, Shandong 264209, China}
\altaffiltext{16}{National Optical Astronomy Observatory, 950 North Cherry Avenue, Tucson, AZ 85719, USA}

\date{Received:~ Accepted:~}
\begin{abstract}

In this paper, we report our high-resolution ($0^{\prime\prime}.20\times0^{\prime\prime}.14$ or $\sim$$70\times49$\,pc) observations of the \midco\ line emission, which probes warm and dense molecular gas, and the 434\,\mum\ dust continuum in the nuclear region of NGC 7130, obtained with the Atacama Large Millimeter Array (ALMA).  The CO line and dust continuum fluxes detected in our ALMA observations are $1230\pm74$\,Jy\,\kms\ and $814\pm52$\,mJy, respectively, which account for 100\% and 51\% of their total respective fluxes.  We find that the \midco\ and dust emissions are generally spatially correlated, but their brightest peaks show an offset of $\sim$70\,pc, suggesting that the gas and dust emissions may start decoupling at this physical scale. The brightest peak of the \midco\ emission does not spatially correspond to the radio continuum peak, which is likely dominated by an Active Galactic Nucleus (AGN). This, together with our additional quantitative analysis, suggests that the heating contribution of the AGN to the \midco\ emission in NGC 7130 is negligible. The \midco\ and the extinction-corrected Pa-$\alpha$ maps display striking differences, suggestive of either a breakdown of the correlation between warm dense gas and star formation at linear scales of $<$100\,pc or a large uncertainty in our extinction correction to the observed Pa-$\alpha$ image. Over a larger scale of $\sim$2.1\,kpc, the double-lobed structure found in the \midco\ emission agrees well with the dust lanes in the optical/near-infrared images. 
\end{abstract}
\keywords{galaxies: active --- galaxies: nuclei --- galaxies: ISM --- galaxies: starburst 
--- galaxies: evolution --- submillimeter: galaxies}

\section{Introduction}
Luminous infrared galaxies (LIRGs; $L_{\rm IR}$[8$-$1000\,$\mu$m]$>10^{11}$\,$L_\odot$), whose space density exceeds that of optically selected starburst and Seyfert galaxies (AGNs) at comparable bolometric luminosity (Soifer et al. 1987), are a mixture of single galaxies, galaxy pairs, interacting systems and advanced mergers. They exhibit enhanced star formation (SF) rates (SFR), usually in their nuclear region, and host a higher fraction of AGN compared to less luminous galaxies (Sanders \& Mirabel 1996). A detailed study of local LIRGs is critical to our understanding of the cosmic evolution of galaxies and AGNs since the co-moving energy density of the Universe at $z\gtrsim1$ is dominated by LIRGs (i.e. Le Flo\'{c}h 2005; Magnelli et al 2009).

Using the \herschel\ (Pilbratt et al. 2010) SPIRE Fourier Transform Spectrometer (FTS; Griffin et al.~2010) data on a flux-limited sample of 125 LIRGs from the Great Observatories All-Sky LIRG Survey (GOALS; Armus et al.~2009),  Lu et al.~(2014) found that SFR correlates much better with the mid-J CO line emission, e.g., CO\,(6-5) or CO\,(7-6), than with the low-J ones ($J\lesssim4$).  This has been further confirmed by Liu et al. (2015) on an expanded sample also including the star-forming regions in nearby normal galaxies, and by our Atacama Large Millimeter Array (ALMA; Wootten \& Thompson 2009) Cycle-0 high-resolution imaging of two LIRGs (NGC 34 and NGC 1614) in CO\,(6$-$5) (Xu et al. 2014, 2015).  Furthermore, Lu et al.~(2014) showed statistically that any heating contribution of an active galactic nucleus (AGN) to the mid-$J$ ($4<J<10$) CO emission is relatively insignificant.  As a result, a mid-J CO emission line, such as \midco, is not only an excellent tracer of SFR, but also an effective probe to the distribution of the warm and dense molecular gas that is intimately related to the on-going SF in the nuclei of LIRGs.

We have initiated a multi-cycle ALMA program to observe representative LIRGs from our FTS sample to map simultaneously the \midco\ line emission (rest-frame frequency $=691.473$\,GHz) and the dust continuum emission at $\sim$434\,\mum\ in the nuclear region of each target, with an ultimate goal of reaching a linear resolution of $\sim$50 pc or less.  This includes our ALMA Cycle-0 observations of NGC\,34 and NGC\,1614 (Xu et al. 2014, 2015),  which represent advanced mergers with a very warm far-infrared (FIR) color (i.e., $f_{\nu}(60\,\mu {\rm m})/f_{\nu}(100\,\mu {\rm m}) \sim 1$), and Cycle-2 observations of NGC\,7130 (IC 5135) presented here.  The latter was from our ALMA Cycle-2 targets that are more representative of typical LIRGs (e.g., $f_{\nu}(60\,\mu {\rm m})/f_{\nu}(100\,\mu {\rm m})\sim 0.6$), covering both compact nuclear core and circumnuclear disk configurations visible in the high-resolution Pa-$\alpha$ images of Alonso-Herrero et al.~(2002, 2006) and whether there is a significant AGN based on the [Ne\,{\scriptsize V}] observation of Petric et al.~(2011).

NGC 7130 is classified as a peculiar spiral of type Sa. Its major and minor diameters (de Vaucouleurs et al. 1991) imply a disk inclination of $\sim$25$^{\rm o}$ (Lu 1998).  At its distance of 72.7\,Mpc (Table \ref{obslog}; Armus et al. 2009), $1^{\prime\prime}$ corresponds to 352\,pc. It has $L_{\rm IR}=10^{11.42}\, L_\odot$ and a moderately warm FIR color of 0.65. It harbors a Seyfert 1.9 nucleus (V\'{e}ron-Cetty \& V\'{e}ron 2006), which might be Compton thick ($N_{\rm H} > 1.5\times10^{24}$\,cm$^{-2}$; Levenson et al. 2005), and a compact  circumnuclear starburst, with a projected effective radius of $\sim$90\,pc (Gonz\'{a}lez Delgado et al. 1998). The CO\,(1-0) and CO\,(2-1) lines observed by Albrecht et al.~(2007) have widths of $\sim$90\,\kms. Thean et al.~(2000) presented a Very Large Array (VLA) radio continuum image at 8.4 GHz, but with a quite elongated beam hinting at a limited dynamic range in terms of surface brightness sensitivity. Bransford et al. (1998) detected extended radio emission with a lower resolution observation.

\section{Observations and Data Reduction}
\begin{deluxetable*}{cccccccc}
\centering
\tablecaption{Basic properties of NGC 7130 and ALMA Observation Log\label{obslog}}
\tablewidth{0pt}
\tabletypesize{\scriptsize}
\tablehead{
\multicolumn{8}{c}{Basic Properties}
}
\startdata
\multirow{2}{*}{Name} & R.A. (J2000)& Decl. (J2000)& Dist. & $cz$ & Morph. & Spectral Type &$\log\,L_{
\rm IR}$\\
&(hh:mm:ss)&(dd:mm:ss)& (Mpc)  & (\kms) &&&($L_\odot$)\\
(1)&(2)&(3)&(4)&(5)&(6)&(7)&(8)\\
\hline\noalign{\smallskip}
NGC 7130&21:48:19.50& $-$34:57:04.7&72.7&4842&Sa\,pec&Sy\,1.9&11.42\\
\hline\noalign{\smallskip}
\multicolumn{8}{c}{ALMA Observation Log}\\
\hline\noalign{\smallskip}
\multirow{2}{*}{SB}&Date&Time\,(UTC)&Configuration&$N_{\rm ant}$&$l_{\rm max}$&$t_{\rm int}$&$T_{\rm sys}$\\
&(yyyy/mm/dd)&&&&(m)&(seconds)&(K)\\
(1)&(2)&(3)&(4)&(5)&(6)&(7)&(8)\\
\hline\noalign{\smallskip}
X87b480\_Xa11&2014/07/26&02:40:41-03:04:14&C34-5&25&650&315&677
\enddata
\tablecomments{For {\bf basic properties}. Col. 1: source name; Cols. 2 and 3: right ascension and declination; Col. 4: distance; Col. 5: Heliocentric velocity from NASA/IPAC extragalactic database (NED); Col. 6: morphology classification from NED; Col. 7: Nuclear activity classification; Col. 8: total infrared luminosity. For {\bf ALMA observation log}. Col. 1: schedule-block number; Cols. 2 and 3: observation date and time; Col. 4: configuration; Col. 5: number of usable antennae; Col. 6: maximum baseline length; Col. 7: on-source integration time; Col. 8: median system temperature.}
\end{deluxetable*}

We observed the central region of NGC 7310 in \midco\ using the Band 9 receivers of ALMA in the time division mode (velocity resolution: $\sim$6.8\,\kms). The four basebands (i.e., “Spectral Windows”; SPWs 0-3) were centered at the sky frequencies of 680.423, 682.178, 676.693, and 678.631\,GHz, respectively, each with a bandwidth of 2\,GHz. The observations were carried out in the relatively extended configuration C34-5 using up to 32 12-m antennae (7 out of which had problematic data; Table \ref{obslog}), with the baselines in the range of $25.8-820.2$\,m{\footnote{\url{https://almascience.nrao.edu/documents-and-tools/cycle-2/alma-technical-handbook}}}. The total on-source integration time was 315 seconds. During the observations, the phase and gain variations were monitored using J2151-3027 and J2056-472 respectively. The error in the flux calibration was estimated to be 7\% 

The data were reduced with CASA 4.3.1. The primary beam is $\sim$$8^{\prime\prime}.5$. However, the maximum recoverable scale is $\sim$$3^{\prime\prime}.1$. The continuum was estimated using data in SPWs 1-3. For the CO\,(6-5) line emission, the cube was generated using the data in SPW-0, which encompasses the CO\,(6-5) emission at the systematic velocity (4842\,\kms; optical) with an effective bandpass of 800\,\kms. The raw images were cleaned using the Briggs weightings, and have nearly identical synthesized beams, with the full width of half maximum (FWHM) of $\sim$$0^{\prime\prime}.20\times0^{\prime\prime}.14$, corresponding to physical scales of 70\,pc$\times$49\,pc, and a position angle (north to east) of $-79^\circ$. The astrometric accuracy of these ALMA observations is better than $0^{\prime\prime}.07$, whereas the relative position accuracy is about $0.5\times$synthesized beam/signal-to-noise ratio{\footnote {\url{https://help.almascience.org/index.php?/Knowledgebase/Article/View/153/6/what-is-the-astrometric-position-accuracy-of-an-alma-observation}}}. Therefore, the relative position accuracies of the peak emission are $\sim$ 2 and 3 mas, for our continuum and integrated line emission maps, respectively. Unless otherwise stated, flux measurements are based on the images after the primary beam correction, whereas all of the figures are produced using the results before the primary beam correction. 

The spectral cubes were binned into channels with a width of $\delta v=13.5$\,\kms. The noise of these channel maps in \midco\ is on the order of 11\,mJy\,beam$^{-1}$. For the continuum, the $1\sigma$ rms noise is 1.3\,mJy\,beam$^{-1}$, and for the \midco\ line emission map, integrated over the barycentric velocity range of $v = 4655.5{\rm -}4953.5$\,\kms, is 1.1 Jy\,beam$^{-1}$\,\kms. All noise measurements were performed on the maps before the primary beam correction.

It is necessary to check whether our results are contaminated by the side-lobe effect since there are three regions in our maps (see Figure \ref{Figmom}) and the southwest (SW) and northeast (NE) regions appear to be symmetric with respect to the central component. From the synthesized beam image we obtained that the maximum side-lobe level of our observation is around 25\%. However, none of the side-lobe peaks at the NE/SW region, and the closest side-lobe is located $\sim$$0\arcsec.8$ ($0\arcsec.4$) north (southeast) to the peak emission of the NE (SW) region (assuming that the PSF peaks at the central region). These distances are $\gtrsim$the diameter of the central region (N-S direction), and thus the NE/SW emission is unlikely significantly contaminated by the side-lobes of the central region. Indeed, for the NE/SW region, there is no obvious enhancement in the ALMA simulated images compared with the observed one.

\section{Results and Discussion}
                                                                    
\subsection{\midco\ Emission}

\begin{figure*}[th]
\centering
\includegraphics[width=0.77\textwidth,bb = 4 136 548 920]{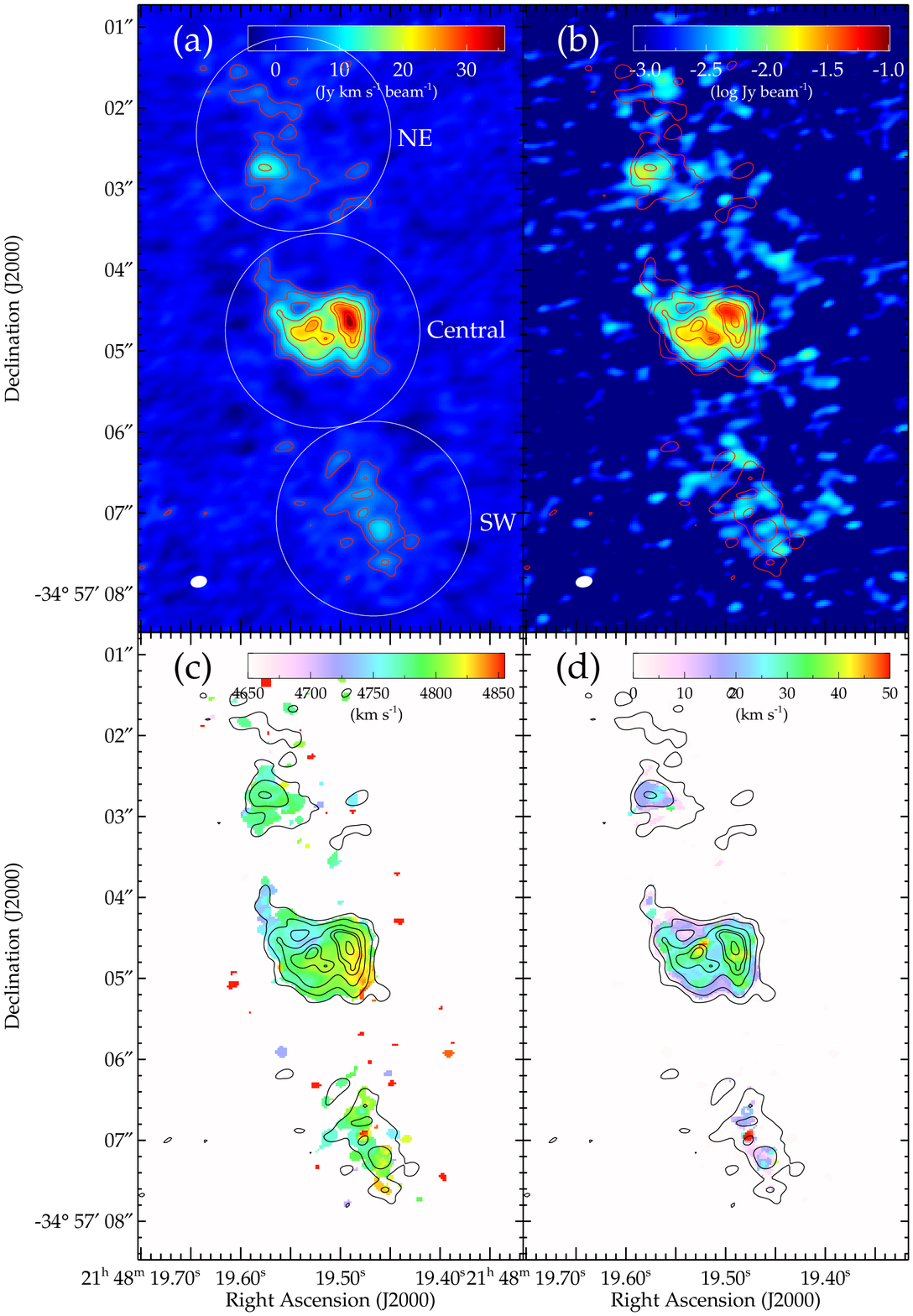}
\caption{CO\,(6-5) line emission contours of the integrated map superimposed on (a) the integrated \midco\ map; (b) the 434\,\mum\ continuum; (c) the first moment map; and (d) the second moment  map. The contour levels are [1, 2, 4, 6, 7, 10]\,$\times$\,$3\sigma$. The beam shapes are shown by the filled (white) ellipse in (a) and (b). The three circles in (a) illustrate the labelled regions used in the text.}
\label{Figmom}
\end{figure*}

Figures \ref{Figmom}a-\ref{Figmom}d show the integrated \midco\ emission, the 434\,\mum\ continuum, the first and second moment maps, respectively. All images are overlaid by the same contours of the integrated \midco\ line emission. In Figure \ref{Figmom}a we marked the 3 distinct regions: NE, central and SW. The central region appears clumpy, and several peaks can be identified, whereas the other two regions look more diffuse. For example, three clumps were found by the clump-finding algorithm of Williams et al. (1994){\footnote{\url{http://www.ifa.hawaii.edu/users/jpw/clumpfind.shtml}}}, which performs a blind search for clumps by contouring the map at different levels to identify peaks. The information of these clumps, such as integrated flux, size and separation, is given in Table \ref{clumpinfo}.

\begin{deluxetable*}{ccccccccc}
\centering
\tablecaption{Information of the three clumps in the central region\label{clumpinfo}}
\tablewidth{0pt}
\tabletypesize{\scriptsize}
\tablehead{\colhead{\multirow{2}{*}{No.}} & \colhead{R.A. (J2000)}& \colhead{Decl. (J2000)}&\colhead{FWHMx} &\colhead{FWHMy} &\colhead{$f_{\rm peak}$} &\colhead{$f_{\rm total}$}& \colhead{Separation}&\colhead{\multirow{2}{*}{Clumps}}\\
&(hh:mm:ss)&(dd:mm:ss)& (\arcsec) &(\arcsec) & (Jy\,\kms) &(Jy\,\kms) &(\arcsec)\\
(1)&(2)&(3)&(4)&(5)&(6)&(7)&(8)&(9)\\
}
\startdata
1&21:48:19.491 & $-$34:57:04.65&0.32&0.51&0.67&196.9&0.58&1,2\\
2&21:48:19.529 & $-$34:57:04.68&0.50&0.53&0.47&196.7&0.28&2,3\\
3&21:48:19.515 & $-$34:57:04.85&0.41&0.34&0.45&120.5&0.42&3,1
\enddata
\tablecomments{Col. 1: clump number; Cols. 2 and 3: right ascension and declination of the clump center; Cols. 4 and 5: FWHM at x- and y-direction, respectively; Col. 6: peak flux; Col. 7: integrated flux; Col. 8: separation between clumps; Col. 9: clump numbers used to calculate the separation listed in column 7.}
\end{deluxetable*}

\begin{figure*}[tbhp]
\centering
\includegraphics[width=0.7\textwidth,bb = 9 15 552 836]{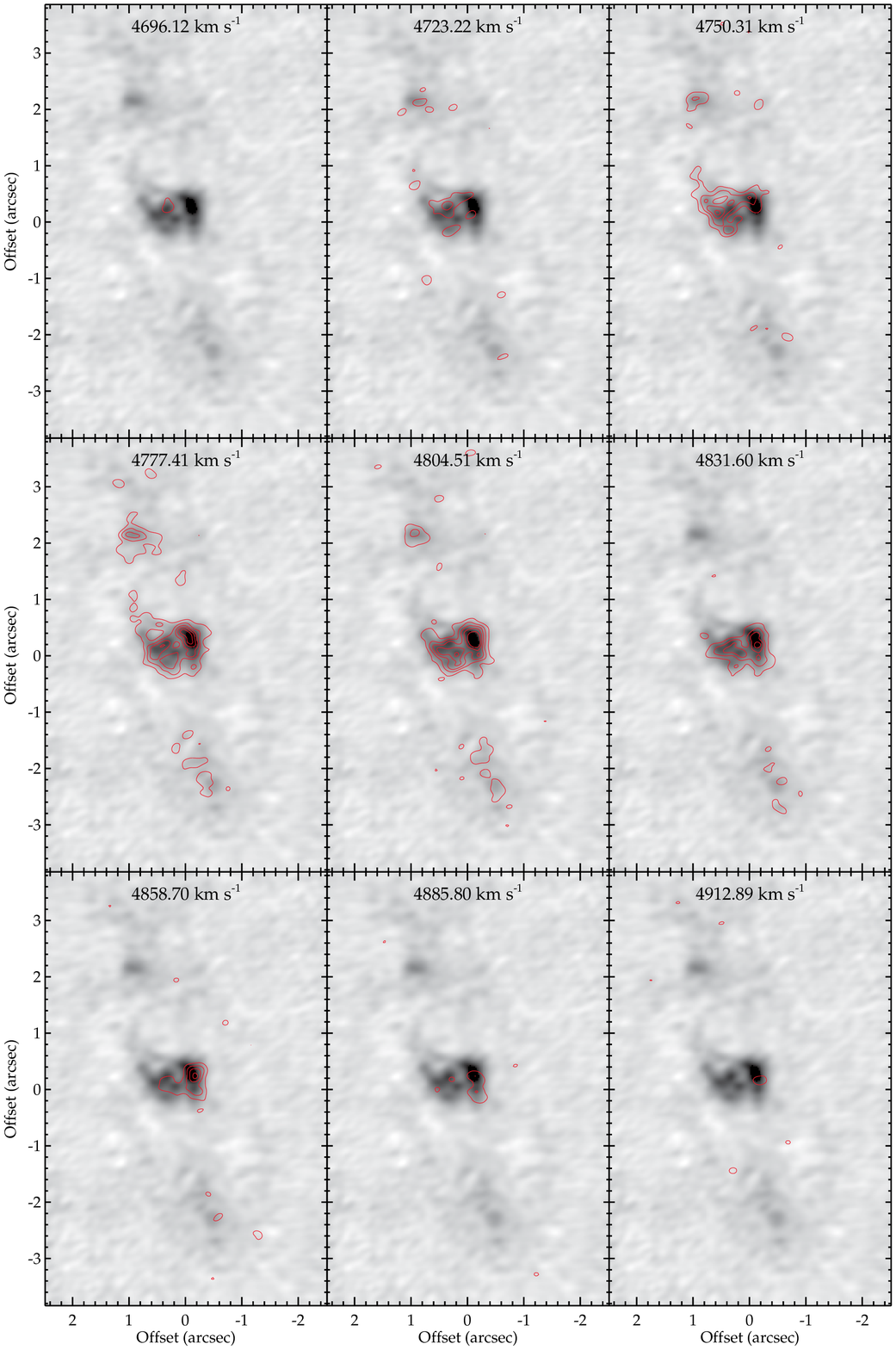}
\caption{CO\,(6-5) line emission contours of the channel maps overlaid on the integrated emission map. The width of each velocity channel is 13.5\,\kms\ and every other channel is displayed. The contour levels are [1, 2, 3, 4, 5]\,$\times$\,$5\sigma\,(\sigma\sim11\,{\rm mJy\,beam}^{-1})$. In each channel, the central barycentric (radio) velocity is labeled.}
\label{Figchannel}
\end{figure*}

The velocity field (Figure \ref{Figmom}c) indicates that the gas in all the 3 regions participates in a common rotation that covers radii up to $\sim$1\,kpc.  The mid point between the two separate CO concentrations in the nucleus has a systemic velocity of $\sim$4790\,\kms.  From this mid point, the underlying (projected) rotational velocity increases radially out, to a value of a few tens of \kms\ seen at the outer edge of the nuclear gas.  At the radii of the NE or SW regions, the underlying rotational velocity is reduced to $\sim$10\,\kms.  This overall kinematic pattern can also been seen in the channel maps displayed in Figure \ref{Figchannel}, where channels are overlaid on the integrated line emission image.  The \midco\ emission in the NE and SW regions coincide well with the dust lane associated with the inner spiral arms as seen in the {\it Hubble Space Telescope} ({\it HST}) optical image of Gonz\'alez Delgado et al.~(1998), suggesting that the observed rotational velocity field likely reflects the galactic rotation.  

Figure \ref{Figmom}d shows that the line-of-sight velocity dispersion is from $\sim$30 to 50\,\kms\ in the central region, around 20\,\kms\ in the NE region, and between 10 and 20\,\kms\ in the SW region. However, the core of the SW region has a large velocity dispersion, as high as $\sim$50\,\kms.

\begin{figure}[tbhp]
\centering
\includegraphics[width=0.47\textwidth,bb = 205 62 630 1015]{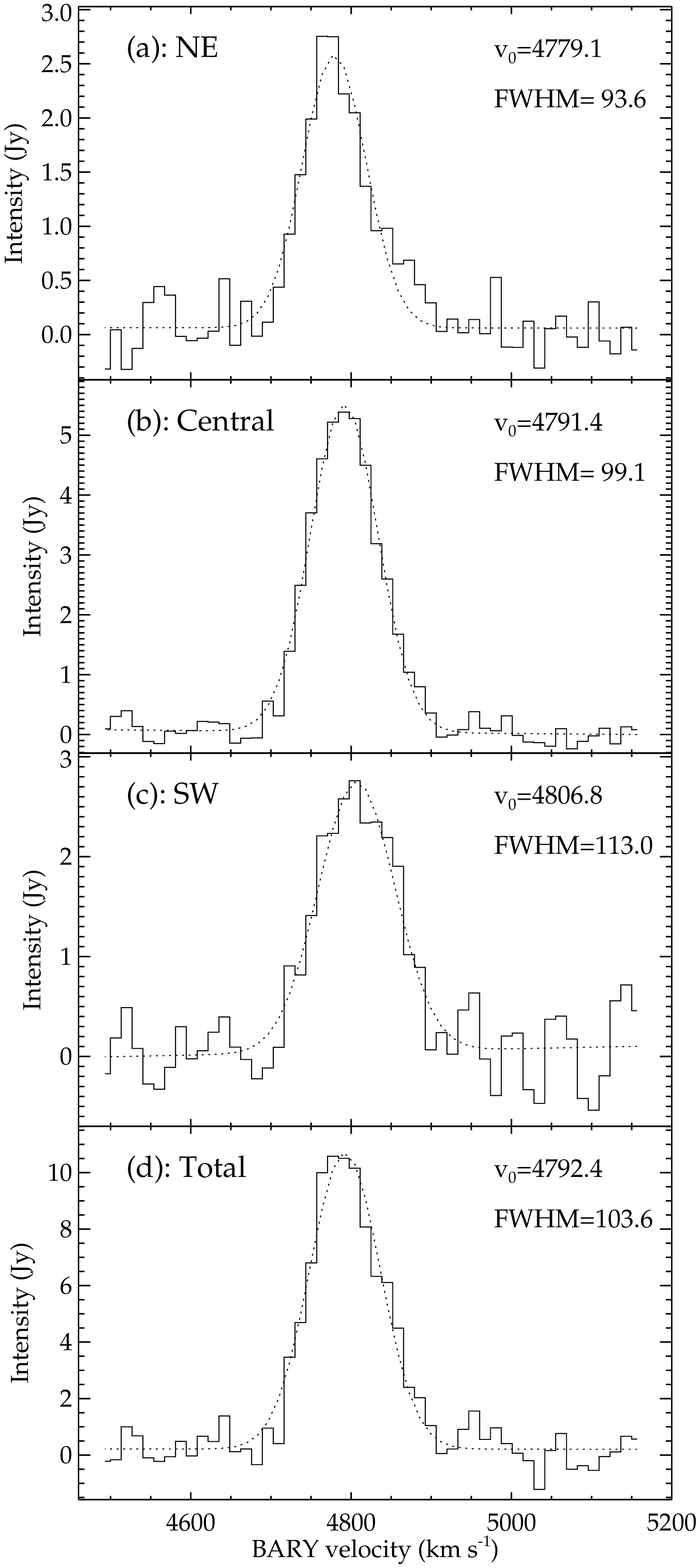}
\caption{Spatially integrated \midco\ line profiles (after primary beam correction) of various regions: (a) NE, (b) Central, (c) SW and (d) the total ALMA detection.  The central velocity and FWHM of a Gaussian fit are given in each plot.}
\label{Figpv}
\end{figure}

Figures \ref{Figpv}a-\ref{Figpv}d present the integrated \midco\ line profiles for the 3 regions as well as for the total ALMA detection, together with the central velocity and FWHM from a Gaussian fit. The central velocities are consistent with a systematic rotation that has a projected velocity of 12 to 15\,\kms\ at the radii of NE or SW.  Among the three regions, the SW one has the broadest line width, consistent with its large velocity dispersion seen in Figure 1d.  The FWHM of the total line emission in Figure~3d is similar to that of CO\,(1-0) and CO\,(2-1) (Albrecht et al.~2007).

The total flux of \midco\ measured from the integrated image is $1230\pm74$\,Jy\,\kms, which agrees well with our SPIRE/FTS-measured flux of $1223\pm82$\,Jy\,\kms\ (Lu et al. 2015, in preparation), obtained with a much larger aperture of $\sim$33$^{\prime\prime}$.   The central ($\sim$$1^{\prime\prime}.5$ in diameter; or $\sim$530\,pc) region contains $\sim$50\% of the total ALMA-detected emission, and the NE and SW regions about 20\% and 30\% respectively. These results suggest that our ALMA observation has recovered all of the \midco\ emission in NGC 7130.

\subsection{Continuum Emission at 434\,\mum}
As shown in Figure \ref{Figmom}b, the 434\,\mum\ continuum emission generally correlates spatially with the \midco\ emission in NGC 7130, but with two notable differences over smaller scales: (1) The brightest peaks of the nuclear continuum and \midco\ line emission have a small offset of $\sim$0$^{\prime\prime}$.2 (or $\sim$70 pc). (2) The average line-to-continuum ratio ($R_{\rm CO/cont}$) is particularly high in the SW region compared to those in the central or NE regions. The corresponding values of $R_{\rm CO/cont}$ are 2100 (SW), 1700 (NE) and 1300 (central), suggesting an intrinsically faint 434\,\mum\ flux density in the SW region.  In star forming regions, the dust emission is always dominated by far-UV photon heating in the photo-dissociation regions (PDRs) near young massive stars. By contrast, the dominant heating mechanism for the \midco\ emission is still controversial despite the fact that its heating source should still be associated with the on-going SF on a global scale (Lu et al. 2014).  The observed differences between the dust continuum and CO\,(6-5) emission at small scales may lend direct support to the notion that shocks (or other mechanical energy input) are mainly responsible for the warm CO emission as often suggested by many authors who modeled the observed CO spectral line energy distributions in galaxies (e.g., Nikola et al.~2011; Rangwala et al.~2011; Kamenetzky et al.~2012; Meijerink et al.~2013; Rosenberg et al.~2014).

This can be better understood in a simplified analytical way: The \midco\ to continuum ratio, $I_{\rm CO\,(6-5)}/(\nu I_{\nu, 434 \mu {\rm m}})$, equals $(I_{\rm CO\,(6-5)}/I_{\rm CO\,(1-0)}) \times (I_{\rm CO\,(1-0)}/(\nu I_{434 \mu {\rm m}}))$. In the inner galaxy where most of the gas is molecular (e.g., Tanaka et al. 2014), $I_{\rm CO\,(1-0)}$ scales with the total gas mass ($M_{\rm gas}$); $I_{\nu, 434 \mu {\rm m}}$ is likely in or near the Rayleigh-Jeans limit and thus scales approximately with the dust mass ($M_{\rm dust}$) times the effective dust temperature ($T_{\rm dust}$).  Therefore, $I_{\rm CO\,(6-5)}/(\nu I_{\nu, 434 \mu m}) \propto (I_{\rm CO\,(6-5)}/I_{\rm CO\,(1-0)})(M_{\rm gas}/M_{\rm dust})(1/T_{\rm dust})$. Since $M_{\rm gas}/M_{\rm dust}$ is more or less fixed, a higher $I_{\rm CO\,(6-5)}/(\nu I_{\nu, 434 \mu {\rm m}})$ ratio can result from a higher CO excitation or/and a lower $T_{\rm dust}$. If both \midco\ and dust emissions are tied to the same PDR mechanism, their emission peaks should coincide with each other spatially.  This is not consistent with our data at scales $\le 70$\,pc. Furthermore, if supernova-driven shocks are responsible for the \midco\ emission, the high $I_{\rm CO\,(6-5)}/(\nu I_{\nu, 434 \mu {\rm m}})$ ratio in the SW region could be associated with the epoch when the O stars (with lifetimes of several times $10^6$ years) from the last starburst episode have burned off, leading to a lower $T_{\rm dust}$. Since supernova progenitors are less massive (down to $\sim$$5\,M_{\odot}$) and have longer lifetimes (up to $\sim$$10^8$ years), the $I_{\rm CO\,(6-5)}/I_{\rm CO\,(1-0)}$ ratio could still remain high, leading to a higher $I_{\rm CO\,(6-5)}/(\nu I_{\nu, 434 \mu {\rm m}})$ ratio.  This picture is overall consistent with the observations that the gas in the SW region displays a complex velocity field and little Pa-$\alpha$ emission (see Fig.~4 below). However, there exists another possibility that the NE/SW gas were excited by bar-induced shocks, which can inhibit star formation when the shock and shear is too strong (e.g. Athanassoula 1992; Reynaud \& Downes 1998).

Our ALMA-detected flux density of the continuum at 434\,\mum\ is $f_{434\,\mu{\rm m,\,ALMA}}=814\pm52$\,mJy. Following a similar procedure (i.e., interpolating the SPIRE measurements at 350 and 500\,\mum\ to obtain the total flux) in Xu et al. (2014, 2015), we estimated about $0.51\pm0.05$ of the total dust continuum is detected by our ALMA observation, which is lower by a factor of about 2 than the interferometer-to-single-dish flux ratio of the \midco\ line emission.  This result indicates that the dust distribution is substantially more extended or more diffuse overall than that of the warm dense gas.

\subsection{Comparison with Other Observations}
\begin{figure*}[t]
\centering
\includegraphics[width=0.75\textwidth,bb = 41 188 688 1152]{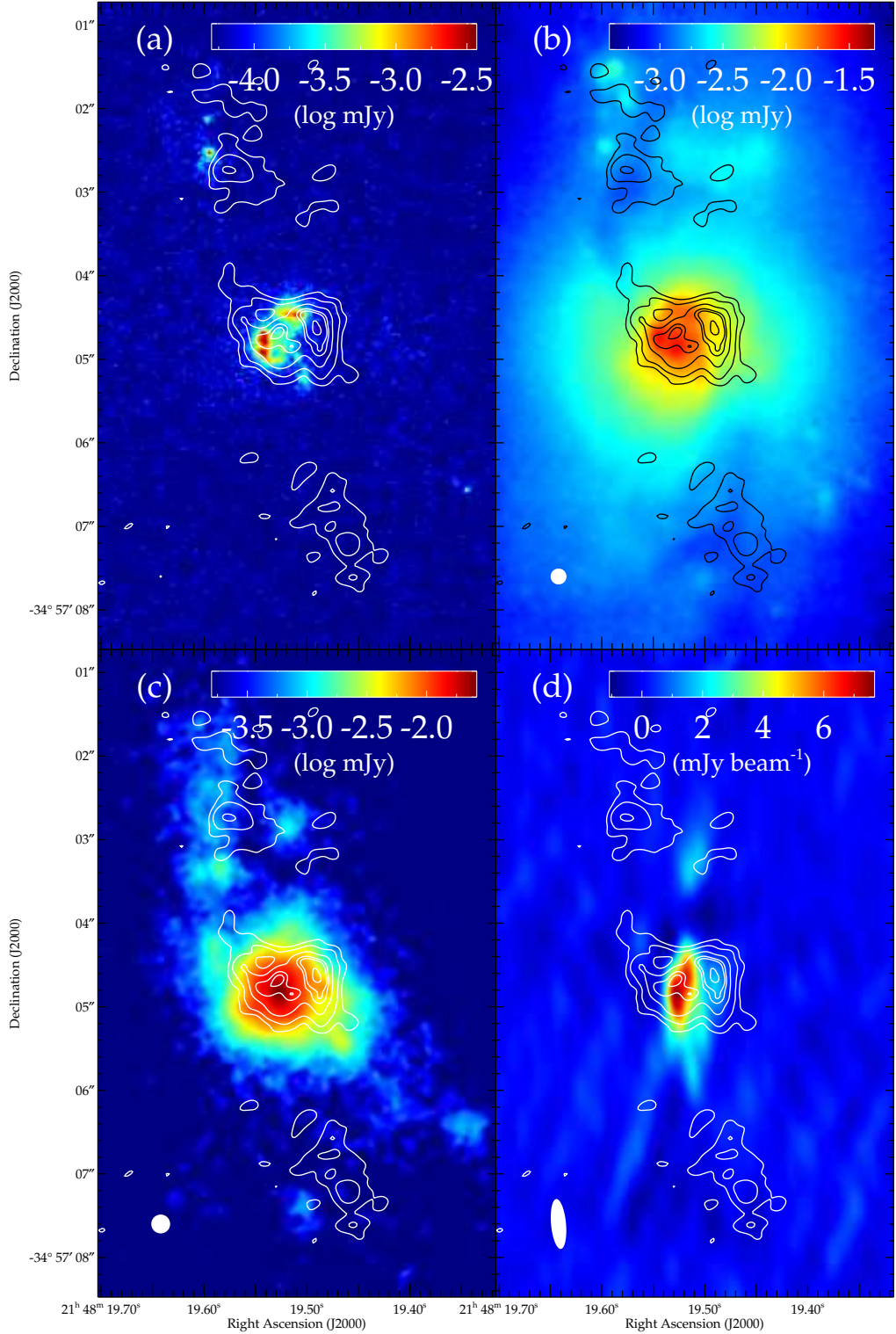}
\caption{Integrated \midco\ line emission contours overlaid on the (a) FOC F210M image; (b) F160W image; (c) extinction-corrected Pa-$\alpha$ image and (d) VLA 8.4\,GHz continuum image. The contour levels are the same as those in Figure \ref{Figmom}. The filled ellipse/circle in the bottom left of each panel illustrates the resolution of the image. The black circle in panel (d) gives the region named A$_{\rm XDR}$.}
\label{Figcomp}
\end{figure*}

Figures \ref{Figcomp}a-d display the integrated \midco\ line emission contours overlaid on the {\it HST} F210M (UV) (Gonz\'alez Delgado et al. 1998), F160W continuum and Pa-$\alpha$ (Alonso-Herrero et al. 2006; Di\'{a}z-Santos et al. 2008), and the VLA 8.4\,GHz continuum images (Thean et al. 2000), respectively. The extinction-corrected Pa-$\alpha$ map was obtained from a set of {\it HST} near-IR (NIR) narrow- and broad-band images (Di\'az-Santos et al. 2008). The radio image was obtained by reducing the raw data downloaded from the NRAO VLA science archive\footnote{\url{https://science.nrao.edu/facilities/vla/archive/index}}. The elongated beam (due to the target being in the southern sky), using a uniform weighting, has a size of $0^{\prime\prime}.60\times0^{\prime\prime}.19$. The maximum recoverable scale for this snapshot observation is about $3^{\prime\prime}.2$, very close to our ALMA observation. 

The astrometry of these {\it HST} observations has been improved by utilizing the 2MASS point source catalog and the ring-like feature identified in Gonz\'alez Delgado et al. (1998). Furthermore, we have also assumed that the nucleus seen in the F160W peak coincides with the radio peak since an AGN usually peaks at the nucleus which is the bottom of the gravitational potential well. Indeed Gonz\'alez Delgado et al. (1998) identified the optical peak (which correlates with the F160W-band) as the nucleus. This way, we can match ALMA and NIR maps to an accuracy of $<0.1\arcsec$, dominated by the ALMA/radio offset.

Figure \ref{Figcomp}a shows that, the clumps in the UV image have offsets relative to the dense gas emission, which may be caused by extinction since the UV data are heavily obscured (Gonz\'alez Delgado et al. 1998), and/or by the break in the local SF law (Xu et al. 2015). The western peak in the \midco\ image seems to be complementary to the ring-like structure shown in the UV image. Figure \ref{Figcomp}b reveals that the two dust lanes, which run in the north-south direction and are also seen in the optical image (Gonz\'alez Delgado et al. 1998), coincide well with our \midco\ emission (SW and NE regions). This indicates that our method for improving the astrometry of the {\it HST} images is reasonable. Furthermore, the dust lanes and NE/SW gas offset towards the leading sides of the inner bar, which can be identified in the $K$-band image and is oriented at ${\rm P.A.}= 0^\circ$ (Mulchaey et al. 1997). This result is consistent with numerical simulations (e.g. Athanassoula 1992) that in the presence of a barred potential, gas will hit the bar from behind and shock, forming a dust-lane, and then the gas will fall inward along the bar. The shocks can heat gas much more efficiently than dust.  Therefore,  both the observed gas morphology and high CO(6-5)-to-continuum ratio indicate that the NE/SW gas might have been accumulated through bar-induced gas inflow.

From Figure \ref{Figcomp}c it is evident that the Pa-$\alpha$ emission varies smoothly with radius, and and has no clumps corresponding to the ones seen in the central region of the \midco\ image. Unless the extinction correction is severely uncertain, our result indicates a breakdown of the correlation between SF and warm dense gas at sub-100\,pc scales, which has been found in NGC 1614 (Xu et al. 2015; but see Wu et al. (2010) and Chen et al. (2015) for opposite results). Nevertheless, a quantitative analysis, using a higher-quality radio image, is needed to reach a solid conclusion since the extinction correction for the Pa-$\alpha$ image may suffer from some uncertainties. 

The only peak seen in the radio continuum image (Figure \ref{Figcomp}d) is close to the eastern clumps of the \midco\ emission (central region). There is no detectable radio emission for the NE and SW regions $\sim$2$^{\prime\prime}$ away from the central region seen in \midco, whereas Liu et al. (2015) found that there is no difference in the global FIR and radio correlations with dense gas and CO emission. The majority of and the peak emission of the radio map may correspond to the AGN, as a compact radio core has been detected in NGC 7130 at 2.3\,GHz (Corbett et al. 2002). This is also supported by the fact that, for star-forming regions in NGC 1614, the average \midco-to-radio flux ratio ($R_{\rm CO/radio}$) is $5.4\times10^4$\,\kms, and for NGC 34, whose AGN contribution to the radio continuum is negligible (Corbett et al. 2002), $R_{\rm CO/radio}=6.1\times10^4$\,\kms. For the western knot (i.e. the western peak of the CO emission in the central region) in the radio map, $R_{\rm CO/radio}$ is $5.3\times10^4$\,\kms. Whereas for the region centered at the radio peak (measured within an ellipse of $a=0^{\prime\prime}.6$ and $b=0^{\prime\prime}.2$) in NGC 7130, it is $1.1\times10^4$\,\kms, $\sim$5 times lower than the forementioned values, indicating a significant enhancement of the radio emission in this region.

Assuming that the X-ray dominated region associated with the AGN is unresolved by ALMA, we measured the \midco\ flux within the region ($r=0^{\prime\prime}.2$; hereafter A$_{\rm XDR}$) centered at the radio peak and found that it only accounts for $\sim$15\% of our ALMA-detected flux within the central region, or $\sim$7\% of the total flux of the entire galaxy. However, there might be contribution from massive stars to the heating of \midco\ in A$_{\rm XDR}$. So it is an upper limit of the AGN contribution to the \midco\ heating. Our results are also consistent with the finding in Lu et al. (2014), i.e., \textit{SF rather than AGN dominates the excitation of mid-J CO emission}. The spectrum extracted within A$_{\rm XDR}$ has a central velocity of 4799\,\kms\ and FWHM of 86.4\,\kms, which are $\sim$7\,\kms\ and 12\,\kms\ larger and smaller, respectively, than those of the spectrum from the entire central region. However, unlike the optical lines from the nucleus (e.g. Davies et al. 2014), no extended wing was detected in all of our \midco\ spectra (see Figure \ref{Figpv}). 

\section{Summary}
In this paper, we presented the initial results from our ALMA band-9 imaging spectroscopy of NGC\,7130 with the following main findings: (a) Half of the \midco\ emission is contained in a clumpy nuclear region of $\sim$500\,pc in diameter and with the rest in two symmetrically placed structures at distances of $\sim$500 pc from the nucleus, coinciding with the optical dust lanes along the inner spiral arms. (b) The \midco-emiting gas appears to participate in the general rotation of the galaxy disk. (c) While the \midco\ emission and the 434\,\mum\ dust continuum are spatially correlated at large scales, there is a $\sim$70 pc offset between their emission peaks, suggesting that the two emissions might become decoupled at this characteristic scale. (d) While the AGN in NGC\,7130 likely dominates the radio continuum emission, the radio continuum has a poor spatial correspondence with the \midco\ emission, confirming our earlier conclusion that AGN heating plays an insignificant role in the mid-J CO line emission.

\begin{acknowledgements}
We thank the anonymous referee for useful comments. Y.Z. and Y.G. acknowledge support by NSFC grants No. 11173059, 11390373 and 11420101002, and CAS pilot-b program \#XDB09000000. This paper makes use of the following ALMA data: ADS/JAO.ALMA\#2013.1.00524.S. ALMA is a partnership of ESO (representing its member states), NSF (USA), and NINS (Japan), together with NRC (Canada) and NSC and ASIAA (Taiwan), in cooperation with the Republic of Chile. The Joint ALMA Observatory is operated by ESO, AUI/NRAO, and NAOJ. This research has made use of the NASA/IPAC Extragalactic Database (NED), which is operated by the Jet Propulsion Laboratory, California Institute of Technology, under contract with the National Aeronautics and Space Administration.
\end{acknowledgements}

\end{document}